\documentclass[twocolumn,aps,showpacs]{revtex4}
\usepackage{graphicx}
\usepackage{dcolumn}
\usepackage{bm}

\begin{document}

\title{Electrically Charged Strange Quark Stars}

\date{\today}

\author{Rodrigo Pican\c{c}o Negreiros}
\email{negreiros@sciences.sdsu.edu} \affiliation{Department of
  Physics, San Diego State University, 5500 Campanile Drive, San
  Diego, CA 92182, USA}

\author{Fridolin Weber} \email{fweber@sciences.sdsu.edu}
\affiliation{Department of Physics, San Diego State University, 5500
  Campanile Drive, San Diego, CA 92182, USA}

\author{Manuel Malheiro}
\email{malheiro@ita.br}
\affiliation{Instituto Tecnol\'{o}gico da A\'{e}ronautica,
 Pra\c{c}a Marechal Eduardo Gomes, 50, Vila das Ac\'{a}cias,
 S\~{a}o Jos\'{e} dos Campos, SP, Brazil}

\author{Vladimir Usov} \email{Vladimir.Usov@weizmann.ac.il}
\affiliation{Center for Astrophysics, Weizmann Institute, Rehovot
  76100, Israel}

\begin{abstract}
  The possible existence of compact stars made of absolutely stable
  strange quark matter--referred to as strange stars--was pointed out
  by E.\ Witten almost a quarter of a century ago.  One of the most
  amazing features of such objects concerns the possible existence of
  ultra-strong electric fields on their surfaces, which, for ordinary
  strange matter, is around $10^{18}$~V/cm.  If strange matter forms a
  color superconductor, as expected for such matter, the strength of
  the electric field may increase to values that exceed
  $10^{19}$~V/cm. The energy density associated with such huge
  electric fields is on the same order of magnitude as the energy
  density of strange matter itself, which, as shown in this paper,
  alters the masses and radii of strange quark stars at the 15\% and
  5\% level, respectively. Such mass increases facilitate the
  interpretation of massive compact stars, with masses of around $2 \,
  M_\odot$, as strange quark stars.
\end{abstract}

\pacs{21.65.Qr; 26.60.Gj; 97.10.Cv; 97.60.Jd}

\maketitle

\section{Introduction}

The possibility that strange quark matter (strange matter for short)
\cite{farhi84:a,alcock86:a,alcock88:a,madsen98:b}, made up of roughly
equal numbers of unconfined up, down and strange quarks, may be the
absolute ground state of the strong interaction is known as the
strange quark matter hypothesis
\cite{bodmer71:a,witten84:a,terazawa89:a}. In the latter event objects
made of strange matter--ranging from strangelets at the low baryon
number end
\cite{farhi84:a,berger87:a,schaffner92:a,gilson93:a,weber93:b,glen94:a,%
  kettner94:b,zhang02:a,alford08:b} to compact stars at the high
baryon number end
\cite{alcock86:a,alcock88:a,madsen98:b,weber05:a,glen97:book,weber99:book,%
  page06:a,sedrakian07:a,golf09:a}--would be more stable than their
non-strange counterparts--atomic nuclei and neutron stars. On
theoretical scale arguments, strange matter is as plausible a ground
state as the confined state of hadrons
\cite{witten84:a,farhi84:a,alcock88:a,glen91:a}. Despite of several
decades of research, both theoretical and experimental, there is no
sound scientific basis on which one could either confirm or reject the
hypothesis so that it remains a serious possibility of fundamental
significance for physics and astrophysics \cite{weber05:a}. One very
striking consequence of the hypothesis is the prediction of the
existence of a new class of compact stars--called strange (quark)
stars \cite{alcock86:a,alcock88:a,madsen98:b,malheiro03:a}. The
heavier members of this hypothetical family of compact stars have
masses and radii similar to those of neutron stars. In contrast to
neutron stars, however, strange stars would form a distinct and
disconnected branch of compact stars and are not part of the continuum
of equilibrium configurations that include white dwarfs and neutron
stars \cite{weber93:b,glen94:a,kettner94:b}.

If strange stars should exist in the Universe, they ought to be made
of chemically equilibrated strange matter, which requires the presence
of electrons inside strange stars. The presence of electrons plays a
crucial role for strange stars, since they may cause the formation of
an electric dipole layer on the surfaces of such stars leading to huge
electric fields on the order of $10^{18}$~V/cm
\cite{alcock86:a,alcock88:a}. The situation is even more extreme if
strange stars were made of color superconducting strange matter
\cite{alford08:a}, which could be either in the color-flavor-locked
(CFL) phase
\cite{rajagopal01:a,alford01:a,linares06:a,rajagopal01:b,alford02:a}
or in the 2-flavor color superconducting (2SC) phase
\cite{rajagopal01:a,alford01:a,page06:review}. In the latter event the
electric fields on the surfaces of quark stars may even be on the
order of $10^{19}$~V/cm \cite{usov04:a,usov05:a}, depending on
electrostatic effects, including Debye screening, and the surface
tension of the interface between vacuum and quark matter
\cite{jaikumar05:a,alford06:a}. The energy density $E^2 / 8\pi$ of
such tremendously large electric fields is on the same order of
magnitude as the energy density of strange quark matter itself and,
thus, should be incorporated in the energy-momentum tensor that is
used to describe strange quark stars. This paper outlines how this is
accomplished mathematically and discusses the consequences of this
extra contribution for the bulk properties of strange stars.

\section{General Relativistic Sellar Structure}

Our discussion is performed for spherically symmetric strange quark
matter stars, whose metric is specified by the line element ($ds^2 =
g_{\nu \mu} dx^\nu dx^\mu$, where $\nu, \mu=0,1,2,3)$,
\begin{eqnarray}
ds^2 = e^{\Phi(r)}c^{2}dt^{2}-e^{\Lambda(r)}dr^{2}-r^{2}(d\theta^{2}
   +\sin^{2}\theta d\phi^{2}) \, . \label{metr}
\end{eqnarray}
The properties of the stellar matter enter Einstein's field equation,
$G_\nu{}^\mu = (8 \pi/c^4) T_\nu{}^\mu $, through the energy-momentum
tensor, $T_{\nu}{}^{\mu}$, which for electrically uncharged matter is
given by
\begin{eqnarray}
  T_{\nu}{}^{\mu} = (P +\rho \, c^2)u_{\nu} u^{\mu} + 
  P \, \delta_{\nu}{}^\mu \, ,
\label{eq:emt1}
\end{eqnarray}
where $P$ is the pressure and $\epsilon = \rho c^2$ the energy density
of strange matter, and $u^{\nu}$ its contravariant four velocity. The
equation of state of strange matter is computed from the MIT bag
model,
\begin{equation}
  P = (\epsilon - 4\, B) / 3 \, , \label{EOS}
\end{equation}
for a bag constant of $B^{1/4} = 150$~MeV. This places the energy of
strange matter at around 870~MeV, well below the energy per particle
of the most stable atomic nucleus, $^{56}{\rm Fe}$, as well as of
infinite nuclear matter \cite{farhi84:a}.  The presence of strong
electric stellar fields, as considered in this paper, renders the
expression of the energy-momentum tensor Eq.\ (\ref{eq:emt1})
significantly more complicated \cite{ray03:a,malheiro04:a},
\begin{eqnarray}
  T_{\nu}{}^{\mu} &=& (P +\rho \, c^2)u_{\nu} u^{\mu} + 
  P \, \delta_{\nu}{}^{ \mu}
  \nonumber \\
  &&+\frac{1}{4\pi} \left( F^{\mu l} F_{\nu l} +\frac{1}{4 \pi}
    \delta_{\nu}{} ^{\mu} F_{kl} F^{kl} \right) \, ,
\end{eqnarray}
where $F^{\nu \mu}$ is the electromagnetic field tensor. The latter
satisfies the covariant Maxwell equations,
\begin{equation} [(-g)^{1/2} F^{\nu \mu}]_{, \mu} = 4\pi j^{\nu}
  (-g)^{1/2} \, ,
  \label{ecem}
\end{equation}
where $j^{\nu}$ stands for the electromagnetic four-current and $g
\equiv \mbox{det}(g^{\nu \mu})$. For static stellar configurations,
which are considered in this paper, the only non-vanishing component
of the four-current is $j^0$.  Because of symmetry reasons, the
four-current is only a function of radial distance, $r$, and all
components of the electromagnetic field tensor vanish, with the
exception of $F^{01}$ and $F^{10}$, which describe the radial
component of the electric field. From Eq.\ (\ref{ecem}) one obtains
the following expression for the electric field,
\begin{equation}
  E(r) = F^{01}(r)=  \frac{1}{r^2} e^{-(\Phi + \Lambda)/2}  4\pi
    \int_{0}^{r}  r'^2 \rho_{ch} e^{ \Lambda /2} dr' \,
  , \label{comp01}
\end{equation}
where $\rho_{ch} = e^{\Phi/2}j^{0}(r)$ represents the electric charge
distribution inside the star.  The electric charge within a sphere of
radius $r$ is given by
\begin{equation}
  Q(r) = 4\pi \int_{0}^{r}  r'^2 \rho_{ch} e^{\Lambda /2}  dr' \, ,  
  \label{Q}
\end{equation}
which can be interpreted as the relativistic version of Gauss' law.
With the aid of Eqs.\ (\ref{ecem}) through (\ref{Q}) the
energy-momentum tensor can be written as
\begin{small}
\begin{equation}
T_{\nu}{}^{\mu} =\left( \begin{array}{cccc}
-\left( \epsilon + \frac{Q^2}{8\pi r^4} \right)  & 0 & 0 & 0 \\
0 & P - \frac{Q^2}{8\pi r^4} & 0 & 0 \\
0 & 0 & P + \frac{Q^2}{8\pi r^4}  & 0 \\
0 & 0 & 0 & P  +\frac{Q^2}{8\pi r^4}
\end{array} \right) , \label{TEMch}
\end{equation}
\end{small}
where the electric charge is connected to the electric field through the
relation ${Q^2(r)}/{8\pi r^4} = {E^{2}(r)}/ {8\pi}$.  Substituting Eq.\
(\ref{TEMch}) into Einstein's field equation leads to
\begin{eqnarray}
  e^{-\Lambda}\left(\frac{1}{r^{2}}-\frac{1}{r} \frac{d\Lambda}{dr}\right)
  -\frac{1}{r^{2}} = - \frac{8\pi G}{c^4}
  \left( \epsilon + \frac{Q^{2}}{8\pi r^4} \right) \, ,  \label{fe1q} \\
  e^{-\Lambda}\left(\frac{1}{r}\frac{d\Phi}{dr}+\frac{1}{r^{2}}
  \right) -\frac{1}{r^{2}}= - \frac{8 \pi G}{c^4} \left( p -
    \frac{Q^{2}}{8\pi r^4} \right) \, . \label{fe2q}
\end{eqnarray}
In analogy to the electrically uncharged case, next we define
\begin{equation}
  e^{-\Lambda}(r) \equiv 1 - \frac{Gm(r)}{rc^2} +\frac{GQ^2(r)}{r^2 c^4} \, .
  \label{nord}
\end{equation}
Equations (\ref{fe1q}) and (\ref{nord}) then lead to
\cite{bekenstein71:a,felice95:a}
\begin{equation}
  \frac{dm}{dr} = \frac{4\pi r^2}{c^{2}} \epsilon
  +\frac{Q}{c^2 r}\frac{dQ}{dr} \, , \label{dmel}
\end{equation}
where $m(r)$ is the gravitational mass contained in a sphere of radius
$r$.  The first term on the right-hand-sight of Eq.\ (\ref{dmel})
comes from the mass-energy of the stellar matter (quark matter in our
case), while the second term on the right-hand-side has its origin in
the mass-energy of the electric field carried the electrically
charged quark star.

The hydrostatic equilibrium equation that determines the global
structure of electrically charged quark stars is obtained by requiring
the conservation of mass-energy, $T_{\nu}{}^\mu;_\mu =0$. This leads
to \cite{bekenstein71:a}
\begin{eqnarray}
  \frac{dP}{dr}  & = & - \frac{2G\left( m +\frac{4\pi r^3}{c^2}
      \left( P - \frac{Q^{2} }{4\pi r^{4} c^{2}} \right) \right)}{c^{2} r^{2}
    \left( 1 - \frac{2Gm}{c^{2} r} + \frac{G Q^{2}}{r^{2} c^{4}} \right)}
 \ (P +\epsilon)\nonumber \\ & & +\frac{Q}{4 \pi r^4}\frac{dQ}{dr} \, .
\label{TOVca}
\end{eqnarray}
The standard stellar structure equation of electrically uncharged
stars, known as the Tolman-Oppenheimer-Volkoff (TOV) equation, is
obtained from Eq.\ (\ref{TOVca}) for $Q \rightarrow 0$. We also note
that if gravity were switched off, Eq.\ (\ref{Q}) leads to
\begin{equation}
  \nabla^2 \mu_e = 4\pi e^2(n_q - n_e) \, ,
\label{eq:TF1}
\end{equation}
with $n_q$ and $n_e$ denoting the electric charge distributions of
quarks and electrons of strange matter, respectively. For a gas of
free electrons one has $n_e = \mu_e^3/3\pi^2$ so that Eq.\
(\ref{eq:TF1}) becomes the Poisson equation
\begin{equation}
  \nabla^2 \mu_e = 4\pi e^2
  ( n_q - \mu_e^3 / 3 \pi^2 ) \, ,
\label{eq:TF2}
\end{equation}
which determines the electron chemical potential for given electric
quark charge densities $n_q$.

\section{Modeling  the Electric Charge Distribution}

Our goal is to explore the influence of huge electric fields on the
bulk properties of strange stars. As already described at the
beginning of this paper, the electric charge distribution that is
causing these huge electric fields on strange stars is located in the
immediate surface regions of such objects. We model this distribution
in terms of a Gaussian which is centralized at the surface of a
strange star. This is accomplished by the following ansatz,
\begin{equation}
  \rho_{ch}(r) = \kappa \, \exp 
  \bigl( - \left( (r - r_g) / b \right)^2 \bigr) \, ,
  \label{chd}
\end{equation}
where $b$ is a charge constant describing the width of the Gaussian,
and $r_g$ is the radial distance at which the charge distribution is
centralized. The quantity $\kappa$ is a normalization constant to be
determined such that
\begin{equation}
  4\pi \int_{-\infty}^{+\infty} \rho_{ch}(r)  r^{2}dr = \sigma \, ,  
  \label{normk}
\end{equation}
where $\sigma$ is a constant proportional to the magnitude of the
electric charge distribution. In flat space-time, $\sigma$ would be
the total electric charge of the system. This is not the case here,
however, as the electric charge $Q$ depends on the metric (see Eq.\
(\ref{Q})). In the latter event the total electric charge of the
system can only be computed self-consistently. Substituting Eq.\
(\ref{chd}) into Eq.\ (\ref{normk}) leads to
\begin{equation}
  8 \pi \kappa = \sigma \left( \sqrt{\pi} b^{3} / 4 +
    r_{g}b^{2} + \sqrt{\pi} r_{g}^{2} b / 2 \right)^{-1}  \,  ,
\label{eq:kappa}
\end{equation}
which establishes a connection between $\kappa$ and $\sigma$.

\section{Results}

\subsection{Parameters and Boundary Conditions}

\begin{figure}
  \includegraphics[scale=0.35]{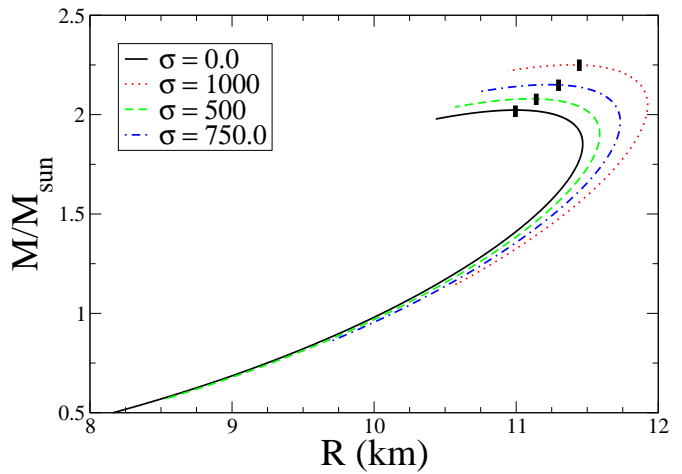}
  \caption{(Color Online) Mass-radius relationship of electrically
    charged strange stars. Tick marks denote the maximum-mass star of
    each sequence, whose properties are given in Table
    \ref{tab:table1}.}
\label{MRfam} 
\end{figure}

Having derived the equations that describe electrically charged
strange stars, we now proceed to solving them numerically and
discussing theirs solutions.  We begin with discussing the boundary
conditions of the TOV equation of electrically charged stars, Eq.\
(\ref{TOVca}). These are: (1) $\epsilon(r=0)=\epsilon_{c}$ which
specifies the energy density at the center of the star, (2) $Q(r=0)=0$
which ensures that the electric charge is zero at the star's center,
(3) $m(r=0)=0$ which ensures that the gravitational mass is zero at
the star's center, and (4) $P(R) = 0$ which defines the surface of the
star located at $r=R$.  Equations (\ref{EOS}), (\ref{Q}),
(\ref{dmel}), (\ref{TOVca}) and (\ref{chd}) can then be solved for the
mass-radius relationship of electrically charged strange stars shown
in Fig.\ \ref{MRfam}, computed for a range of different values for
$\sigma$.  The curve for which $\sigma =0$ describes electrically
uncharged strange stars. For numerical reasons
a sample value of 0.001~km was chosen for the width $b$ of the
electric charge distribution on the surface of a strange star. This
value is many orders of magnitude greater than the true width of the
electric charge distribution, which is only around $\sim 10^3$~fm.
This particular choice, however, has no influence on the numerical
outcome, which can be seen by inspecting
 \begin{equation}
   \frac{dQ}{dr} = \frac{r^{2} \sigma \, \exp
     \bigl( - ((r - r_g)/b)^2 \bigr)  \exp ({\Lambda/2 })}
   {2 (\sqrt{\pi} b^{3} /4 + r_{g}b^{2} + \sqrt{\pi} r_{g}^{2} b / 2)  } \, ,
\label{dqcomp}
\end{equation}
which follows from Eqs.\ (\ref{Q}), (\ref{chd}), and (\ref{eq:kappa}).
Equation (\ref{dqcomp}) reveals that $dQ/dr \rightarrow \sigma
(r/r_g)^2 \exp(\Lambda/2) \delta(r-r_g)$ as the width goes to zero, $b
\rightarrow 0$. Hence, the $dQ/dr$ term in the TOV equation
(\ref{TOVca}) is independent of the width of the electric charge
distribution, provided it is sufficiently narrowly distributed as is
the case for strange stars.

\subsection{Sequences of Electrically Charged Strange Stars}

Figure \ref{MRfam} shows the mass-radius relationship of electrically
charged strange quark stars for different values of the electric
charge constant $\sigma$. The masses and radii of such stars change at
the 10 to 15\% level, depending on the amount of electric charge
carried by the star. Table \ref{tab:table1} lists the properties of
the maximum-mass star of each sequence displayed in Fig.\ \ref{MRfam}.
It is important to note that the electric field strengths listed in
\begin{table}[h]
  \caption{\label{tab:table1} Properties of electrically charged 
    maximum-mass strange quark stars. The quantities $R$ and $M$ denote 
    their radii and gravitational masses, respectively. The stars carry 
    given electric charges, $Q$, which give rise to electric stellar 
    surface fields $E$.}
\begin{ruledtabular}
\begin{tabular}{ccccc}
$\sigma$ & $R$~ (km)& $M$~ ($M_{\odot}$) &$Q$~ ($\times 10^{17}C$) &E~
($10^{19}$ V/cm)\\
\hline
  0 &  10.99  & 2.02 & 0    & 0   \\
500  &  11.1  & 2.07 & 989  & 7.1 \\
750  &  11.2  & 2.15 & 1486 & 10.5 \\
1000  &  11.4  & 2.25 & 1982 & 13.5 \\
\end{tabular}
\end{ruledtabular}
\end{table}
Table \ref{tab:table1} are the electric fields at the surface of the
quark star (where $P \rightarrow 0$). The electron layer outside the
quark star, however, nullifies the electric field, and an observer at
infinity will thus not be able to detect it.

\begin{figure}
  \includegraphics[scale=0.67]{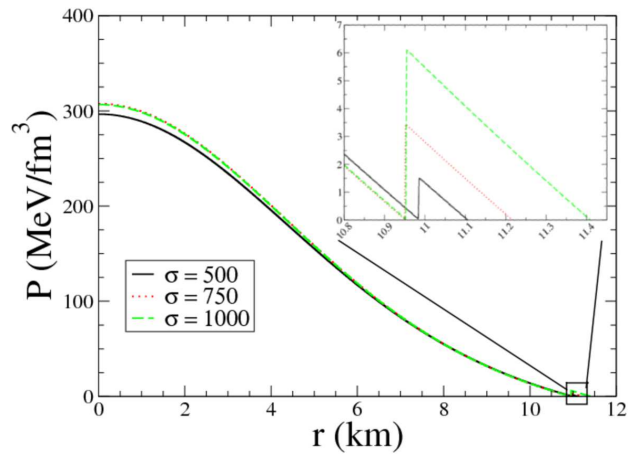}
  \caption{(Color online) Pressure profiles of electrically charged
    maximum-mass strange stars shown in Fig.\ \ref{MRfam}.}
\label{pprof} 
\end{figure}

\begin{figure}
  \includegraphics[scale=0.67]{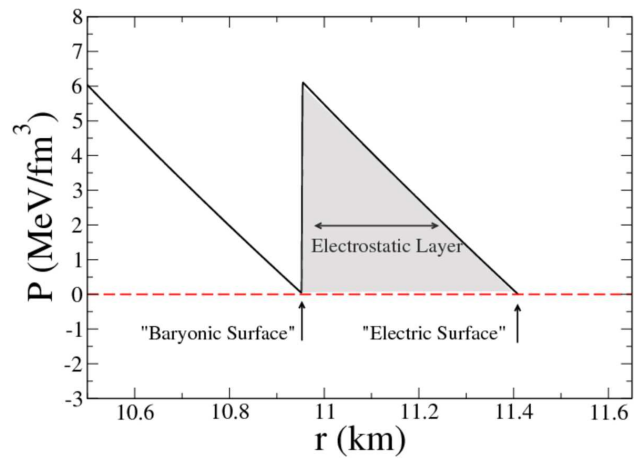}
  \caption{(Color Online) Pressure profile for maximum-mass star for
    the $\sigma = 1000$ sequence. We show the definitions of
    ``Baryonic Pressure'' (where the baryon pressure goes to zero),
    ``Electric Pressure'' where the total pressure of the star
    vanishes, and the electrostatic layer between them.}
\label{p_scheme}
\end{figure}

The pressure profiles of electrically charged strange quark stars
shown in Fig.\ \ref{pprof} exhibit several peculiar features which are
absent in compact stars made of ordinary neutron star matter
\cite{glen97:book,weber99:book,sedrakian07:a}. First, we note that the
interior pressure profiles of quark stars are completely unaffected by
the electric charge layer, since the latter is located in a thin,
spherical shell near the surface of quark stars. As a consequence, the
pressure decreases monotonically from the stellar center toward the
surface, as it is the case for electrically uncharged quark stars and
ordinary neutron stars. The situation changes drastically in the
stellar surface region, however, where the electric charge
distribution causes a sudden and very sharp increase in pressure, as
shown in the insert of Fig.\ \ref{pprof}. This sudden increase in
pressure is a result of the appearance of the ultra-high electric
fields and the drastic change in the total electric charge ($dQ/dr$ is
very high in this region) of the system, resulting in a significant
positive contribution to $dP/dr$ (see Eq.\ (\ref{TOVca})). This
additional pressure, added to the system in a region that would
otherwise be the surface if the star were uncharged, creates a new and
qualitatively different region that surrounds any electrically charged
strange quark star. Electrically charged quark stars thus possess two
surfaces: a ``baryonic matter surface'', where the surface of the
uncharged star would be located, and an ``electric surface'' where the
total pressure of the star vanishes. In between these surfaces is the
ultra-high electric field region, which we will call from now on the
``electrostatic layer'', since it only exists because of the electric
field. These surfaces along with the electrostatic layer are shown in
Fig.\ \ref{p_scheme} for the sample quark star with $\sigma = 1000$
listed in Table \ref{tab:table1}.  The mass increase of the stars
shown in Fig.\ \ref{MRfam} and Table \ref{tab:table1} originates from
the mass-energy added to these stars by the Coulomb field.

\begin{figure}
  \includegraphics[scale=0.30]{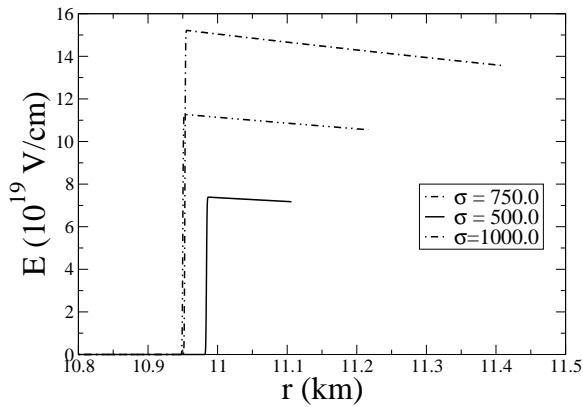}
  \caption{Electric field profile inside maximum-mass stars (plotted
    in the relevant region).}
\label{Eprof}
\end{figure}

Next, we discuss the electric fields of charged quark stars. The
radial dependence of these fields is shown in Fig. \ref{Eprof} for
the maximum-mass stars of Table \ref{tab:table1}. As expected,
the electric fields exhibit a very steep increase at the interface
between the baryonic surface and the electrostatic layer. Since the
electric charge is located in a thin spherical layer, the electric
fields quickly weaken with increasing radial distance, and the
pressure stemming from the charge contributions drops down to zero at
the electric surface of the star. The electric fields are as high as
$10^{19-20}$~V/cm. As already mentioned above, however, an observer at
infinity would not be able to see these electric fields since the star
is surrounded by a layer of electrons that nullifies the electric
fields. An observer outside of the electric layer would thus not be
able to detect it. 

\section{Conclusions}

Almost all of the ambient conditions that characterize compact stars
tend to be extreme. This would specifically be the case for the
electric fields carried by hypothetical compact stars made of
absolutely stable strange quark matter (strange quark stars), which
could be as high as $10^{19}$ to $10^{20}$~V/cm. In this paper, we
perform a detailed investigation of the physical implications of such
ultra-strong fields for the bulk properties of strange quark stars.
The sources of these electric fields are electric charge distributions
located on the surfaces of strange quark stars.  Depending on whether
these stars are made of regular (i.e.\ non-superconducting) strange
quark matter or color superconducting strange quark matter, the
respective charge distributions possess very different physical
characteristics. Color-flavor-locked (CFL) matter, for instance, is
rigorously electrically neutral in bulk
\cite{rajagopal01:b,oertel08:a}.  The situation is
different for strange stars made of regular quark matter, which do
possess electrons in bulk. For such stars two different electric
layers would be formed \cite{alcock86:a}, a positively charged region
of strange quark matter in a thin surface layer of the quark star and
an electron layer outside of the star. The core of such stars has a
net positive charge and the forces acting on it are non-zero. It is
this scenario that can be described by the model presented in this
paper.  Our study shows that the electric charge distribution can have
a significant impact on the structure of quark stars. For this to
happen, the pressure (or equivalently the energy density) associated
with the electric field needs to amount at least a few percent of the
typical pressure $P \sim 100 \text{MeV/fm}^3$ that exists inside of
quark stars. Assuming that $E^2 / 8\pi \sim$ a few $\times 10^{-2}\,
P$ we thus estimate the required electric field strengths as $E \sim
10^{19-20}$~V/cm.  We have shown that electric fields of this
magnitude, generated by charge distributions located near the surfaces
of strange quark stars, increase the stellar mass by up to 15\% and
the radius by up to 5\%, depending on the strength of the electric
field.  These changes are caused by both the sudden increase of
pressure in the surface regions of electrically charged quark stars,
as well as by the energy density of the electric surface field which
acts as an additional energy-momentum source in relativistic
gravity. They facilitate the interpretation of massive compact stars,
with masses of around $2 \, M_\odot$, as strange quark stars.

Last but not least, another interesting feature that we have
discovered is that the electric charge gradient term, $dQ/dr$ (Eq.\
(\ref{dqcomp})), which emerges in the TOV equation (\ref{TOVca}) of
any electrically charged compact star, is independent of the width of
the electric charge distribution, provided the charge distribution is
sufficiently narrowly spread over the star. This finding is not
limited to strange quark stars but applies to any general relativistic
stellar object that carries a narrowly concentrated electric charge
distribution.

\section*{Acknowledgments}

This work was supported by the National Science Foundation under Grant
PHY-55873A ARRA. Manuel Malheiro acknowledges financial support from
CNPq and FAPESP.




\end{document}